\documentclass[prl,twocolumn,tightlines,showpacs,preprintnumbers,aps,floatfix,epsf,amsmath,amssymb]{revtex4}
\usepackage{epsfig,amsmath,amssymb,bm,epsf,graphics}
\usepackage{dcolumn}

\newcommand{\be}{\begin{equation}}
\newcommand{\ee}{\end{equation}}
\newcommand{\ben}{\begin{eqnarray}}
\newcommand{\een}{\end{eqnarray}}
\newcommand{\bF}{\begin{figure}}
\newcommand{\eF}{\end{figure}}
\newcommand{\dg}{\dagger}

\def\ket#1{ | #1 \rangle}

\begin{document}


\title{ SuperSolid Phase in Helium-4 }
\author{Mukesh Tiwari}
\email{mukesh@unm.edu} 
\affiliation{Consortium of the Americas for Interdisciplinary Science, Department of Physics and Astronomy, University of New Mexico, Albuquerque, New Mexico 87131-1156, USA.  }
\author{ Animesh Datta}
\email{animesh@unm.edu}
\affiliation{Department of Physics and Astronomy, University of New Mexico, Albuquerque, New Mexico 87131-1156, USA.}

\date{\today}

\begin{abstract}
The probable observation of a supersolid  helium phase was recently reported by Kim {\it et al}. In this article, we confirm their speculation. Based on a theoretical model for solid helium-4, we show that the emergence of such a phase is inevitable. This is the first instance of Bose-Einstein condensation (BEC) in a solid. We calculate the BEC transition temperature ($T_c$) and the Landau critical velocity ($v_c$) from the model, which respectively are 215 mK and $251 \mu \rm{ms^{-1}}$. They are in excellent agreement with the experimental results; $T_c=$175 mK and $v_c \leq 300  \mu \rm{ms^{-1}}$. We also prove that our model possesses the necessary and sufficient condition for the emergence of supersolidity. We briefly comment about similar behaviour in $^3$He.

\end{abstract}



\maketitle


	Helium has been one of the most exotic elements that has aroused an immense amount of theoretical and experimental interest for almost a century \cite{hist}. Recently another possible first in the realm of Helium, the supersolid phase was reported \cite{kc}. This is believed to the first example of a Bose Einstein condensation (BEC) in solids. The speculations about the existence of such a phase has been around for more than 30 years \cite{leggett},\cite{chester}, \cite{AL} together with numerous attempts to experimentally confirm them \cite{meisel}.

	As is described later, helium is a quantum solid. Vacancies and interstitials are essentially delocalized due to their ability to tunnel through potential barriers. At low temperatures these point defects form a weakly interacting Bose gas. Also large zero point motions result in exceptionally rapid exchange between helium atoms at nearby sites. BEC of either of these two channels may provide for  a new phase of matter with long range crystalline and supersolid order coexisting. This is the supersolid phase. According to Andreev, Lifshitz and Chester \cite{AL}, \cite{chester}, condensation of the vacancies and defects in solid Helium should result in a supersolid phase. It has however, been argued \cite{guyer} that the vacancies in $^4$He are too sparse to result in any meaningful phase transition. On the other hand, Leggett's proposal \cite{leggett} is tantamount to saying that supersolidity will result due to tunneling of real He atoms between neighbouring sites in the crystal. According to estimates based on his predictions, the supersolid fraction should  be $\sim  10^{-6}$ (Leggett's own estimate was $\sim 3 \times 10^{-4}$) and the transition temperature $\ll$ 1 mK.
 	    
	The results of the experiment \cite{kc} are, however, quite different. It is found that the transition occurs at 175 mK and the supersolid fraction is $5 - 25 \times 10^{-3}$. We present here an argument that explains the emergence of the supersolid phase in He with the right parameters. The theory, developed in \cite{gp} is peculiar to solid $^4$He. It is a model to treat the local motion of atoms in solid $^4$He.These modes are suugested to play a role in mass transport usually attributed to point defects (vacancies). The theory has also been successful in resolving the controversy regarding the contribution of point defects to the specific heat of solid $^4$He. From that we are also able to conclude that the helium in Vycor glass is essentially in a bcc phase and not in an hcp phase. We also make contact with the concept of off-diagonal long range order (ODLRO) that is a necessary and sufficient condition for superfluidity in helium \cite{yang}.

\begin{figure}
\begin{center}
\centerline{\epsfig{file=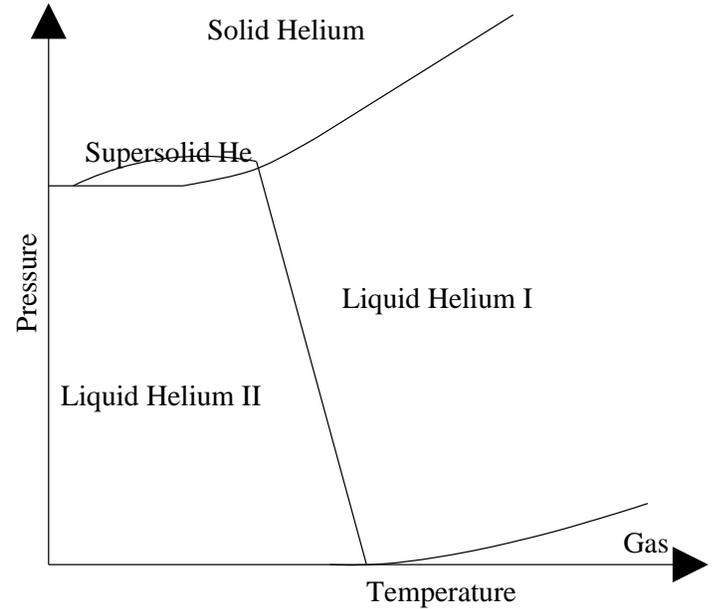,height=3.2in,angle=0}}
\narrowtext{\caption {Schematic Phase diagram of $^4$He.}}
\end{center}
\end{figure}


	We next  review the essential points in the theoretical description of solid $^4$He as propounded in Ref. \cite{gp}. The usual assumption concerning atomic motion in a solid is that the atom resides at a local minimum of the interatomic potential. In the quantum mechanical description however there is always a residual energy $E_z$ even at zero temperature (zero point motion) which is well estimated by $E_z \approx \frac{\hbar^2}{m a^2}$, where $m$ is the mass of the atom and $a$ is the localization of the ground state. This makes itself most evident at the lowest densities. For Helium, $E_z$ is so large that it remains a liquid under its vapour pressure down to absolute zero. Substantial pressure is needed to solidify it, as seen from Fig 1. At low temperatures, the minimal pressure necessary is 25 bar for $^4$He and somewhat more so for $^3$He. Much more pressure (40 bar) is needed to solidify helium in confined geometies \cite{AUTH}. It is perhaps the only solid whose phase diagram is so strongly influenced by quantum mechanics. The subsequent discourse is valid only for low density solids ($\sim 21$ cm$^3$ molar volume). 

	Within a small distance $\Delta x$ around the minimum, the interatomic potential $V(x)$ can be Taylor expanded as 
\be
V(x) \approxeq  V_0 + \frac{1}{2}\Bigg|\frac{\partial^2 V}{\partial x^2}\Bigg|_{x=x_0}(\Delta x)^2  + \dots
\ee
where $V_0$ is the minimum value and $x_0$ is the minima in the potential. However, for the solid helium atom potential  there is a problem: the second derivative at the centre of the well is negative. A related problem in the conventional description is that the zero point motion is too large for the potential to be considered stationary.

	 A key assumption of the harmonic treatment is that the nucleus is a point-like object as compared to the electron cloud. This approximation (Born-Oppenheimer approximation) is not valid when the potential is anharmonic. In a quasi-classical picture, the nucleus will not remain at the centre due to the forces from the anharmonic potential. This relative displacement gives rise to a dipole moment. Dipolar fluctuations of an atom in free space are random and isotropic. In a solid, the motion-induced dipole moments have preferred orientations. Over time, the dipole-dipole interaction between these dipole moments would average to zero if the oscillation frequency and phase of each of the atoms is random. If however, the zero-point motions are correlated, then the time averaged interaction is non-zero and can in certain situations lower the energy of the solid. The lowest interaction energy is evidently when all the dipoles in a lattice are oscillating in phase. Since the direction of the dipole shows the instantaneous direction of the motion, such a state is a translation of the system and hence unphysical. Looking for other symmetric arrangements, one finds two "antiferroelectric" configurations along the symmetry axes of the crystal with individual dipole moments oriented along the (001) direction, as depicted in Fig. 2 in the bcc lattice \footnote{The triangular geometry in a hcp lattice does not allow the type of ordering discussed.  Thus there is no long range order in terms of dipole-moments of the atoms at the lattice points. The uncorrelated state is the minimum energy configuration.} These have a total dipole moment of zero. Similarly, correlated motion can occur along the (010) and (100) axes too. It should however be noted that only one of these will exist in a region of space in the crystal to preserve this long-range ordering. 

\begin{figure}
\begin{center}
\epsfig{file=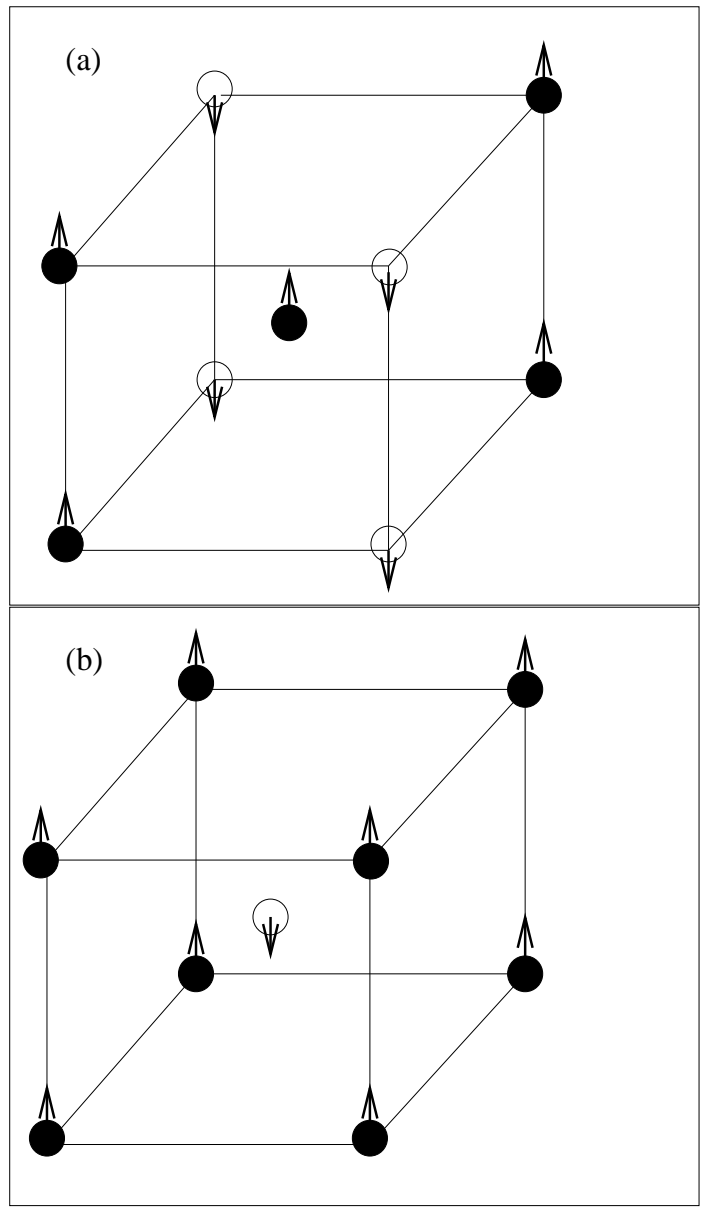,height=6.3in,width=5.3in,angle=0}
\narrowtext{\caption {The two "antiferroelectric" arrangements in the bcc phase. Adapted from \cite{gp}.}}
\end{center}
\end{figure}

	The dipoles introduced above are affected by lattice vibrations, viz, phonons. For a modulation along a direction 
$\bm{k}$, the dipolar interaction energy is 
\be
X(\bm{k})= -|\mu|^2 \sum_{i\neq0}\Bigglb[\frac{3\cos^2[\bm{\mu}\cdot(\bm{r}_0-\bm{r}_i)]-1}{|\bm{r}_0-\bm{r_i}|^3}\Biggrb]e^{2 \pi i \bm{k}\cdot(\bm{r}_0-\bm{r}_i)},
\ee
where $\bm{\mu}$ is the induced dipole moment and $\bm{r}_i$ is the instantaneous position of the $i$ th atom. 
	In the bcc geometry, the only phonon modes that modulate the energy along the (001) direction are $L(001)$, $T(100)$ and $T_1(110)$. It can be shown from periodicity properties that the coupling of the local modes to the lattice excitations is only through the $T_1(110)$ mode \cite{gp}. Thus the only elementary excitations to the dipole array would be in the (110) direction, in the form of the $T_1(110)$ phonon.

	 For further analysis one introduces the Hamiltonian for the local interactions which are treated as bosons. The net Hamiltonian written in swcond quantized notation is 
\be
H = \sum_k [\epsilon_0 + X(k)]\Biglb(b^{\dg}_k b_k + \frac{1}{2}\Bigrb) + \sum_k X(k)(b^{\dg}_k b^{\dg}_{-k} + b_k b_{-k})
\label{ham}
\ee 
where $b^{\dg}_k,b_k$ are the bosonic creation/annihilation operators. This Hamiltonian can be diagonalised using the Bogoliubov
transformation $\beta_k = u(k)b_k + v(k) b^{\dg}_{-k}$. The functions $u(k), v(k)$, given in \cite{gp} are not essential to our analysis.The energy spectrum of the diagonalised Hamiltonian is 
\be
\epsilon(k)=\sqrt{\epsilon_0[\epsilon_0+2X(k)]}.
\ee
Since the mode is an acoustic phonon, it must be gapless as $k\rightarrow 0$ and hence $X(k=0)=\epsilon_0$. Using this
\be
\epsilon(k)= \epsilon_0\sqrt{2}\sin[\bm{k}\cdot((\bm{r}_0-\bm{r}_i)]= \epsilon_0\sqrt{2} \sin ka,
\ee
in which the last inequality is particular to the $T_1(110)$ mode.

     As has been indicated in \cite{gp}, this mode can result in ``phonon-assisted'' tunnelling of the actual atoms between the lattice sites. In that case, the atoms are in a 'free-flowing' state that can undergo Bose Einstein condensation and subsequently lead to supersolidity. This is essentially the mechanism suggested by Leggett \cite{leggett}. The difference is that we no more use the Hartree approximation to calculate the dispersion relation of these modes as given in \cite{guyer}; we rather account for the anharmonicity through local modes to obtain the actual coherent diffusion process. 

\vskip0.4cm	
	One can calculate the supersolid transition temperature from the elementary theory of non-interacting Bose gases. The average  number of particles (bosons) is given as 
\be
N = \langle n_{\epsilon}\rangle = \int_0^{\infty} \frac{g(\epsilon)d \epsilon}{z^{-1}e^{\beta \epsilon} -1}
\ee
 where $g(\epsilon)$ is the density of states. For a dispersion relation given as 
\be
\epsilon(k) = \epsilon'_0 \sin ka
\ee
putting $\epsilon'_0=\sqrt{2} \epsilon_0$
\be
g(\epsilon) = \frac{V}{a^3 2 \pi^2 \epsilon'_0} \frac{(\sin^{-1} \epsilon/\epsilon'_0)^2}{\sqrt{1-(\epsilon/\epsilon'_0)^2}}
\ee
with $V$ being the specific volume ($\sim$ 21cc/mole). The lattice spacing in solid helium is $a$ = 2.6 \AA. It immediately follows that
\be
\label{tc}
N = \frac{V}{a^3 2 \pi^2 \epsilon'_0}\int_0^{\epsilon'_0}\frac{(\sin^{-1} \epsilon/\epsilon'_0)^2}{z^{-1}e^{\beta \epsilon} -1}\frac{1}{\sqrt{1-(\epsilon/\epsilon'_0)^2}} d\epsilon.
\ee 
	Since $N$ is a monotonically increasing function of $z$, the condensation takes place at $z=1$. Also the Landau critical velocity is given by the minimum value of ${\frac{\epsilon(p)}{p}}$, where $p = \hbar k$. Thus
\be
\label{lanvel}
v_c = \min \Bigl\{\frac{\epsilon(p)}{p}\Bigr\} = \frac{a \epsilon'_0}{\hbar}\frac{\sin x_0}{x_0}
\ee
$x_0$ being a non-trivial root of $\tan x = x$. Finally, a judicial estimate of  the superfluid fraction is given by
\be
\label{sfrac}
\frac{\rho_s}{\rho} = \epsilon_0/E_z = \epsilon'_0(\hbar^2/m a^2)^{-1}.
\ee

A value of $\epsilon'_0 \sim 30 \mu$K or $\epsilon_0 \sim 30/\sqrt2 = 21.32 \mu$K reproduces the results of the experiment \cite{kc} to excellent agreement. This value is also in harmonious agreement with the data from neutron scattering experiments, $\epsilon_0 \sim 20 \mu$K presented in \cite{guyer}. There the magnitude of the coherent diffusion in solid helium is estimated. On the other hand, with the experimental value of $\epsilon_0 = 20 \mu$K, for a supersolid fraction of $10 \times 10^{-3}$, we obtain from (\ref{tc}) a transition temperature of 215 mK and from (\ref{lanvel}) a critical velocity of $251 \mu \rm{m s^{-1}}$. The transition temperature compares well with the experimentally determined value of 175 mK. The critical velocity is within the low temperature limit of $300 \mu \rm{ms^{-1}}$.   The introduction of the channel predicted in \cite{gp} accounts for the atomic self-diffusion in a way that is in commensuration with other facts like the specific heat of solid helium \cite{leg}. Both these processes however call for the long range correlations characteristic of a quantum solid.

\vskip0.4cm
 	The existence of off-diagonal long range order (ODLRO) is a necessary and sufficient condition for the existence of superfluidity \cite{kohn}. The charecterization of supersolidity used in \cite{kc} is the most literal one, taken from superfluidity: the incomplete dragging by slowly rotating walls within which the material is confined. This is self-evident from their experimental setup. The proof in \cite{kohn} about the necessity  and sufficiency of ODLRO for superfluidity can be trivially extended to the case of supersolidity \cite{kohnnote}. Hence we can conclude that a system is supersolid or superfluid if and only if it exhibits ODLRO. Which of these will occur is of course determined by the thermodynamic coordinates of the system in the phase space, i.e, its temperature, pressure and density. 

        From the Hamiltonian analysis previously carried out \cite{gp}, one finds the ground state wave function of the local modes to be
\be
\ket{\Psi_0} = \prod_k \exp{\Bigl(\frac{v_k}{u_k} b^{\dg}_{k}b^{\dg}_{-k}\Bigr)}\ket{{\rm vac}}.
\ee
where $u_k, v_k$ are the coefficients of the Bogoliubov transformation made in diagonalising the hamiltonian in Eqn. \ref{ham}. 
This, as is well-known, is the famous Bogoliubov wave-function and known to exhibit ODLRO \cite{yang}. The Bose-Einstein condensation is the simplest form of ODLRO and we believe that the emergence of the supersolid phase in solid $^4$He is nothing but the BEC of the $T_1(110)$ phonon mode. Therefore the existence of ODLRO in the local mode ground state wave function reaffirms the role of ODLRO in supersolid phase transition.

\vskip0.4cm
	In the present article we have showm that the emergence of a supersolid phase is a natural consequence of the quantum  nature of solid $^4$He. Although such a phase was proposed a long time ago, the precise physical mechanism for such transition was only recently presented. Experimental efforts in the mean time continued. We can now conclude that the supersolid transition in solid $^4$He is due to a Bose-Einstein condensation of the actual mass transport process amongst the helium atoms at different sites in the bcc phase. The dispersion relation for this process is given by that of the $T_1(110)$ phonon mode (or any of the two other orthogonal ones). For helium solidifying in the hcp phase, there is no long range correlation (not to be confused with ODLRO) between the atomic dipoles and hence no phonon mode analogous to the one dealt with in this paper. Thus $^4$He solidifying in the hcp phase will not exhibit a supersolid transition. For the experiment of solidifying helium in Vycor,as in Ref. \cite{kc} it might be tough to determine directly into which phase, bcc or hcp the helium crystallizes. However the amount of helium going into the bcc/hcp phase can be determined from the value the supersolid fraction. As per our estimates, almost all of it goes into the bcc phase. We also show that the system is an ideal candidate for exhibiting a supersolid transition as it has ODLRO.

	As to $^3$He, as similar behaviour may be expected to exist due to the existence of an identical mode \cite{gp1}. The arguments leading to this mode are however valid only at temperatures much higher than the magnetic interactions (T $\gg 1$mK). $^3$He is a fermion and the superfluid phases in the liquid state are observed at around 2 mK and presumably due to the formation of Cooper pairs. The exact temperature of transition can be hard to calculate due to the interactions in $^3$He \cite{dml}. Nonetheless, if such a supersolid phase is to exist it must be at temperatures at least 2 orders of magnitude below those for $^4$He. Indeed pure $^3$He shows no such transition at temperatures of $\approx 175$mK \cite{kc}.

	We would like to thank M. de Llano for helpful discussions.

\end{document}